# ATTOSECOND PULSE AMPLIFICATION IN A PLASMA-BASED X-RAY LASER DRESSED BY AN INFRARED LASER FIELD

V.A. Antonov[1,2,*], K.Ch. Han[3], T.R. Akhmedzhanov[3], Marlan Scully[3] and Olga Kocharovskaya[3]

[1]Institute of Applied Physics of the Russian Academy of Sciences, Nizhny Novgorod, Russia

[2]Prokhorov General Physics Institute of the Russian Academy of Sciences, Moscow, Russia

[3]Department of Physics and Astronomy, Texas A&M University, College Station, USA

We suggest a technique to amplify a train of attosecond pulses, produced by high-harmonic generation (HHG) of an infrared (IR) laser field, in an active medium of a plasma-based X-ray laser. This technique is based on modulation of transition frequency of the X-ray laser by the same IR field, as used to generate the harmonics, via linear Stark effect, which results in redistribution of the resonant gain and simultaneous amplification of a wide set of harmonics in the incident field. We propose an experimental implementation of the suggested technique in active medium of $C^{5+}$ ions at wavelength 3.4 *nm* in the "water window" range and show the possibility to amplify by two orders of magnitude a train of attosecond pulses with pulse duration down to 100 *as*. We show also a possibility to isolate a single attosecond pulse from the incident attosecond pulse train during its amplification in optically deep modulated medium.

Ultrashort coherent X-ray pulses provide a unique combination of record-high spatial and temporal resolution, which finds numerous applications, ranging from dynamical imaging of nanostructured materials and controlling chemical reactions to manipulation of absorption and ionization properties of atoms on a sub-fs time scale [1-20]. One of the most promising applications of such pulses is an ultrafast imaging of large biological molecules, in particular, proteins. It requires sub-fs pulses containing relatively large number of photons with a carrier frequency in the so-called water window (4.4-2.3 nm– between K-shell absorption edges of carbon and oxygen). The pulses with down to tens of attosecond duration can be produced via high harmonic generation (HHG) of an IR laser field [19-28]. However, in the "water window" range these pulses contain a relatively small number of photons, corresponding to the energies not exceeding few pJ (although much higher energies of sub-fs waveforms can be achieved at longer wavelengths using different schemes [29-32]).

In this work, we suggest a technique for amplification of a train of attosecond pulses, produced by HHG, in active medium of a plasma-based X-ray laser [33-42] in the presence of a strong co-propagating IR laser field. An amplification of a single high-order harmonic by the plasma-based X-ray laser has been widely studied before [36-38, 40-42]. But narrow linewidths of X-ray lasers did not allow for joint amplification of several harmonics. However, in the presence of a strong IR field, the frequency of transition of the X-ray laser becomes modulated with the period equal to a half-cycle of the optical field [43]. As shown below, in such a case the active medium amplifies X-ray radiation not only at the resonant frequency, but also at its multiple sidebands separated by twice the frequency of the modulating field.

Suppose the train of X-ray pulses has the carrier frequency equal to the time-averaged frequency of the modulated transition and its spectrum consists of a set of harmonics separated by doubled frequency of the modulating field. Experimentally, such a situation could be achieved if replicas of the same IR field are used for HHG and for modulation of the active medium. In this scenario, various spectral components of the incident X-ray field might be jointly amplified during their propagation through the medium. In the case of sufficiently dense plasma, considered in this paper, the length of coherent interaction between the IR and X-ray fields is much shorter than the amplification length of the X-ray field, which effectively eliminates scattering of the harmonics into each other. As shown below, with the proper choice of parameters of the mod-ulating field (its amplitude and frequency), it allows amplifying attosecond pulses preserving their spectral and temporal structure.

Let us consider amplification of a set of high-order harmonics in an active medium of a plasma-based X-ray laser with a population inversion at the transition between the ground and the first excited energy levels of hydrogen-like ions, $n=1\leftrightarrow n=2$ (where $n$ is the principal quantum number), simultaneously irradiated by the co-propagating IR laser field of the fundamental frequency. If the harmonic of order $2k+1$ is tuned in resonance with the inverted transition, that is $\omega_{2k+1} = \bar{\omega}_{tr}$, then the frequency of any other harmonic of order $2(k+l)+1$ can be represented as

$$\omega_{2(k+l)+1} = \bar{\omega}_{tr} + 2l\Omega, \tag{1}$$

where $\Omega$ is the laser frequency, $l$ is an integer number, $\bar{\omega}_{tr} = \frac{3}{8}\frac{m_e e^4}{\hbar^2}Z^2\left(1 - \frac{109}{Z^6}\frac{E_C^2}{E_A^2}\right)$ is the frequency of the resonant transition accounting for the quadratic Stark effect [44], $E_C$ is the laser field strength, $E_A = m_e^2 e^5/\hbar^4 \cong 5.14\cdot 10^9 V/cm$ is the atomic unit of electric field, $e$ and $m_e$ are charge and mass of electron, respectively, $\hbar$ is Planck's constant, and $Z$ is nucleus charge number of the ions. Based on Eq. (1), for the sake of conciseness we will call harmonic order $2(k+l)+1$ as "$2l$-th" harmonic.

In order to gain an insight into the process of high-harmonic amplification, we derive an analytical solution for an output X-ray field assuming that the population difference at the resonant transition is constant, so that the amplification of the harmonics occurs in linear regime. We also assume that the IR pulse duration used for both HHG and modulation of the active medium is sufficiently long, so that it can be represented as $\vec{E}_{IR}(x,t) = \vec{z}_0 E_C \cos(\Omega t - x n_{pl}/c)$, where $x$-axis is the propagation direction, $\vec{z}_0$ is a unit polarization vector along $z$-axis, $c$ is the speed of light in vacuum, and $n_{pl}$ is the plasma refractive index for the IR field. The incident X-ray field, comprising the set of harmonics ranging from $2(k-l_{min})+1$ to $2(k+l_{max})+1$, for the sake of analytical study is presented in the form:

$$\vec{E}_{X-ray}(x\leq 0,\tau) = \frac{1}{2}\vec{z}_0 \sum_{l=-l_{min}}^{l_{max}} E_{inc}^{[2(k+l)+1]} \exp\{-i(\bar{\omega}_{tr} + 2l\Omega)\tau\} + \text{c.c.}, \tag{2}$$

where $E_{inc}^{[2(k+l)+1]}$ is a complex amplitude of harmonic, $\tau = t - x/c$ is the local time, and c.c. stands for complex conjugation. Since the harmonics have the same $z$-polarization as the modulating IR field, amplification of the harmonics can be described within the three-level approximation, taking into account the two excited states dressed by an IR field: $|2\rangle=(|2s\rangle+|2p,m=0\rangle)/\sqrt{2}$, and $|3\rangle=(|2s\rangle-|2p,m=0\rangle)/\sqrt{2}$, as well as the ground state $|1\rangle=|1s\rangle$ [43, 45]. If the population differences between the states $|1\rangle$, $|2\rangle$ and $|3\rangle$ remain constant and a phase shift acquired by the IR field due to plasma dispersion at the length of the medium is much larger than π, then (as it is discussed in Supplemental Materials, see Eqs. (S1)-(S12)) the harmonics scattering into each other is strongly suppressed so that each harmonic propagates through the medium independently from the others, and the output X-ray field acquires a simple form

$$\vec{E}_{X-ray}(x,\tau) = \frac{1}{2}\vec{z}_0 \sum_{l=-l_{min}}^{l_{max}} E_{inc}^{[2(k+l)+1]} \exp\{g_{total} J_{2l}^2(p_\omega)x\}\exp\{-i(\bar{\omega}_{tr} + 2l\Omega)\tau\} + \text{c.c.}, \tag{3}$$

where $g_{total} = \frac{4\pi n_{tr} N_{ion} d_{tr}^2 \bar{\omega}_{tr}}{\hbar \bar{\gamma}_{tr} c}$ is the amplification coefficient for the resonant X-ray field in the absence of linear Stark effect, $n_{tr} = \rho_{22} - \rho_{11} = \rho_{33} - \rho_{11}$ is population difference at the transitions $|2\rangle\leftrightarrow|1\rangle$ and $|3\rangle\leftrightarrow|1\rangle$, $d_{tr}$ and $\bar{\gamma}_{tr}$ are the absolute value of dipole moment and the decoherence rate at these transitions, $N_{ion}$ is density of the resonant ions, $J_m(x)$ is Bessel function of the first kind of order $m$, $p_\omega = \Delta_\omega/\Omega$ is the modulation index, and $\Delta_\omega = 3\dfrac{m_e e^4}{\hbar^2 Z}\dfrac{E_C}{E_A}$ is amplitude of linear Stark shift induced by the modulating optical field.

As follows from Eq. (3), irradiation of the active medium of X-ray laser by the modulating field results in appearance of a gain for the X-ray field at the frequencies of harmonics, $\bar{\omega}_{tr} + 2l\Omega$, $l = \pm 1, \pm 2, ...$, at the cost of reduced gain at the central frequency $\bar{\omega}_{tr}$, see Fig. 1(a),(b). Since (i) each spectral component of the incident X-ray field is amplified independently from the others, (ii) the resonant interaction with the ions does not change the phases of harmonics, and (iii) the plasma dispersion for the X-ray field is negligible, the relative phases of harmonics remain constant during propagation through the medium. Thus, if each harmonic experiences the same gain, then the incident field (2) will preserve its temporal shape during the amplification.

The gain for "$2l$-th" harmonic is proportional to the squared Bessel function $J_{2l}^2(p_\omega)$ of order $2l$ of the modulation index $p_\omega$. As follows from Fig. 1(c), the magnitudes of several Bessel functions of even orders are approximately equal at some particular values of the modulation index. For example, for $p_\omega = 6.4$ one has $J_0^2(p_\omega) \simeq 0.059$, $J_2^2(p_\omega) \simeq 0.090$, $J_4^2(p_\omega) \simeq 0.087$, and $J_6^2(p_\omega) \simeq 0.084$. Thus, the IR field providing such value of the modulation index allows for nearly uniform amplification of "0-th", "$\pm 2$-nd", "$\pm 4$-th", and "$\pm 6$-th" harmonics. A number of harmonics which can be amplified increases with increasing $p_\omega$ and is approximately equal to $p_\omega + 1$. Particularly, the values $p_\omega = 10$, $p_\omega = 13.4$, $p_\omega = 16.1$, and $p_\omega = 19.4$ are suitable for amplification of 11, 15, 17, and 21 high-order harmonics of the modulating field, respectively. However, with increasing value of the modulation index the gain, $g_{total} J_{2l}^2(p_\omega)$, averaged over the harmonic order, $2l$, decreases as $1/p_\omega$, while the differences between the amplification coefficients of neighboring harmonics grow (see Fig. 1(c) and Fig. S1, Fig. S6 in Supplemental Materials).

The analytical theory of the linear amplification regime in the framework of the three-level model allows understanding some general aspects of the high-harmonic amplification. However, in order to perform a quantitative analysis, one needs to take into account two additional degenerate upper states $|4\rangle = |2p, m=1\rangle$ and $|5\rangle = |2p, m=-1\rangle$, leading to generation of $y$-polarized amplified spontaneous emission (ASE), and variation of the population differences at all the involved transitions. This more general 5-level nonlinear model is described in Supplemental Materials (see Eqs. (S13)-(S16) and corresponding discussion). Here we present the results of calculations. Let us consider neutral plasma consisting of $C^{5+}$ ions, electrons, and some other ions (for example, $H^+$, to maintain electric neutrality) with $C^{5+}$ ion density $N_{ion} = 10^{19} \, cm^{-3}$ and electron density $N_{el} = 15 N_{ion}$. Lasing in inverted plasma with such parameters has being theoretically studied [39, 46] and is under experimental investigation in the group of Prof. Szymon Suckewer at Princeton University, USA. As the modulating field and a source of the incident high-harmonic signal let us consider 2.1 $\mu m$ mid-infrared laser radiation, which is particularly suitable for HHG in the "water window" [24]. Let us study the case $p_\omega = 6.4$ and consider amplification of the harmonics of this laser field with orders ranging from 617 to 629, which are "0-th", "$\pm 2$-nd", "$\pm 4$-ht", and "$\pm 6$-th" harmonics with respect to the resonant transition. The value $p_\omega = 6.4$ corresponds to intensity of the modulating field $I_C = 2.7 \times 10^{15} \, W/cm^2$. In order to tune "0-th" harmonic (of order 623) in exact resonance with the transitions $|2\rangle, |3\rangle \leftrightarrow |1\rangle$, the laser wavelength should be $\lambda_C = 2102.9 \, nm$. For the numerical study, an incident X-ray field is assumed to be a train of attosecond pulses with Gaussian envelope centered at $t_{peak}$ and the duration (the full width at half maximum of intensity) $t_{1/2}$:

$$E_z(t, x=0) = \frac{1}{2}\vec{z}_0 E_{hh} \exp\left\{-2\ln 2\left(t - t_{peak}\right)^2 / t_{1/2}^2\right\} \sum_{l=-l_{max}}^{l=l_{max}} \exp\left\{-i\left(\bar{\omega}_{tr} + 2l\Omega\right)t\right\} + \text{c.c.}, \quad (4)$$

where $l_{max}=3$. Eq. (4) implies that the incident harmonics are phase synchronized and have identical amplitudes. As initial conditions, we assume that at $\tau = 0$ all the ions are excited to the states $|2\rangle$-$|5\rangle$ with equal probability, while spontaneous emission at the inverted transitions is taken into account via the randomly distributed along the medium initial values of quantum coherencies following the approach developed by Gross and Haroche [47] (see the Supplemental Materials).

The time-dependencies of intensities of (i) z-polarized amplified attosecond pulse train and (ii) y-polarized ASE at the output from a plasma channel with length $L = 1\,mm$ and radius $R = 1\,\mu m$ are shown in Fig. 2. The panels (a), (b), and (c) correspond to different peak intensities, $I_0 = \frac{c}{8\pi}(2l_{max}+1)^2 E_{hh}^2$, of the incident field, namely $I_0 = 10^{13}\,W/cm^2$ (a), $I_0 = 10^{12}\,W/cm^2$ (b), and $I_0 = 10^{11}\,W/cm^2$ (c). Fig. 2 is plotted for $t_{peak} = 10\,fs$ and $t_{1/2} = 35\,fs$, which means that the incident attosecond pulse train (4) reaches the peak amplitude $10\,fs$ after creation of the population inversion by a pumping laser pulse at $\tau = 0$. In such a case, the incident z-polarized X-ray field has a nonzero value of the slowly-varying amplitude at $\tau = 0$, which facilitates its amplification before the development of y-polarized ASE (the influence of $t_{peak}$ on the amplification process is discussed in Supplemental Materials, see Figs. S2-S4). The role of ASE depends on the peak intensity of the incident X-ray field, $I_0$. If the incident z-polarized field is strong enough, Fig. 2(a), then it is amplified and saturates the resonant transitions $|2\rangle\leftrightarrow|1\rangle$ and $|3\rangle\leftrightarrow|1\rangle$ before y-polarized ASE becomes substantial. As a result, population from the sates $|2\rangle$ and $|3\rangle$ drops down to the state $|1\rangle$, reducing population differences at the transitions $|4\rangle\leftrightarrow|1\rangle$ and $|5\rangle\leftrightarrow|1\rangle$ and decreasing amplification of ASE. For this reason, in Fig. 2(a) ASE is negligible. However, as intensity of the incident X-ray field decreases (see Fig. 3(b),(c)), its amplification takes place in a linear regime (without saturation of the resonant transitions), resulting in (i) larger ratio of the output intensity to $I_0$, and (ii) stronger ASE, which in this case saturates the transitions $|4\rangle\leftrightarrow|1\rangle$ and $|5\rangle\leftrightarrow|1\rangle$ and thereby reduces population differences at the transitions $|2\rangle\leftrightarrow|1\rangle$ and $|3\rangle\leftrightarrow|1\rangle$ (see Supplemental Materials, Fig. S5, for more details).

Amplification of shorter pulses is possible with stronger modulating fields. An increase in intensity of the modulating field to $I_C = 2.5 \times 10^{16}\,W/cm^2$ allows amplification of 140 $as$ pulses, produced from 21 high-order harmonics of the laser field with 2.1 $\mu m$ wavelength (in such a case $p_\omega = 19.4$). But the gain decreases with increasing value of the modulation index. Thus, for the same parameters of the medium as in Fig. 2 and $I_0 = 10^{12}\,W/cm^2$ the incident X-ray field (4) will be amplified only 4.2 times (see Supplemental Materials, Fig. S6, for details). However, if the laser frequency, $\Omega$, is increased proportionally to the laser field strength, $E_C$, then the index of modulation is constant, and the amplification will be more efficient. This case is illustrated by Fig. 3, which corresponds to amplification of seven harmonics (of orders 231-243) of the laser field with wavelength $\lambda_C = 801.53\,nm$. The modulating intensity is $I_C = 1.9 \times 10^{16}\,W/cm^2$ and corresponds to $p_\omega = 6.4$. The parameters of the incident X-ray field are the same as in Fig. 2, except for higher $\Omega$; $I_0 = 10^{13}\,W/cm^2$. If the length of the active medium would be also the same, the result of amplification would resemble that in Fig. 2. However, in Fig. 3 the medium length is increased to $L = 7\,mm$. It results in very large optical depth of the medium for the harmonics, $g_{total} J_0^2(p_\omega) L = 22.7$, and thus, in the strong shortening of the attosecond pulse train due to predominant amplification of its front edge, which extracts the major part of the energy, stored in the population inversion of the medium (see Supplemental Materials, Figs. S7-S9, for the details). For the considered parameters of the medium, the envelope of the amplified attosecond pulse train becomes shorter than the repetition period of the pulses in the train. As a result, the active medium isolates a single pulse with 130 as duration, which is amplified to $I_{max} = 261 I_0$.

The lower limit for the duration of attosecond pulses which can be amplified is set by the upper limit for the intensity of the modulating field, which is determined by the threshold of ionization of the resonant ions from the upper lasing states. For the case of $C^{5+}$ ions the acceptable peak intensity of the modulating field can be estimated as $I_C = 3.5 \times 10^{16} W/cm^2$ and corresponds to the ionization time from the states $|2\rangle$ and $|3\rangle$, $\tau_{ion}^{(2),(3)} = 60 fs$. Scaling of the results, shown in Fig. 3, for this intensity of the modulating field and $\lambda_C \approx 590 nm$ leads to duration of the amplified isolated attosecond pulse $\tau_{pulse} \approx 100 as$.

In conclusion, we have shown the possibility to amplify a set of high-order harmonics of an IR laser field in active medium of a hydrogen-like plasma-based X-ray laser, dressed by a replica of the same field as used for HHG. The amplification occurs due to frequency modulation of the lasing transition by the IR field via the linear Stark effect, which results in redistribution of the gain to the combinational frequencies separated from the frequency of the resonance by even number of frequencies of the IR field. For the specific intensities of the modulating field, nearly uniform gain may be provided for the whole set of harmonics. In sufficiently dense plasma, the harmonics are amplified independently from each other, so that their relative phases remain constant. Thus, if the incident X-ray field represents an attosecond pulse train, it will keep this form during the amplification. We suggest an experimental implementation of this method in active medium of $C^{5+}$ ions and show the possibility to amplify by two orders of magnitude the attosecond pulses with duration down to 100 $as$ at the carrier wavelength 3.4 $nm$ in the "water window" range. In optically deep medium, the duration of the amplified attosecond pulse train is reduced due to predominant amplification of its front edge. If the optical depth of the medium is high enough, the active medium selects a single attosecond pulse, which is amplified much stronger than the other pulses from the train. The amplification of a set of harmonics does not rely on specific phases of sidebands of the resonant polarization of the medium emerging under the action of the modulating field. Thus, in principle, it might be implemented in arbitrary (not only hydrogen-like) active medium, if the depth of frequency modulation of the resonant transition(s) exceeds the frequency of the modulating field.

We acknowledge support from Russian Foundation for Basic Research (RFBR, Grant No.18-02-00924), as well as support from National Science Foundation (NSF, Grant No. PHY-150-64-67), AFOSR (Grant No. FA9550-18-1-0141) and ONR (Grant No. N00014-16-1-3054). The derivation of the analytical solution was supported by the Ministry of Education and Science of the Russian Federation under contract No.14.W03.31.0032 executed at the Institute of Applied Physics of the Russian Academy of Sciences. We deeply appreciate stimulating discussions with Szymon Suckewer, Mikhail Ryabikin and Ilias Khairulin. Texas A&M High Performance Research Computing Center is acknowledged for awarding us the supercomputer time. V.A.A. acknowledges support by the Foundation for the Advancement of Theoretical Physics and Mathematics BASIS.

*Corresponding author: antonov@appl.sci-nnov.ru

[1] F. de Groot, *High-Resolution X-ray Emission and X-ray Absorption Spectroscopy*, Chem. Rev. **101**, 1779 (2001).
[2] J.-F. Adam, J.-P. Moy, *Table-top water window transmission x-ray microscopy: Review of the key issues, and conceptual design of an instrument for biology*, Rev. Sci. Instrum. **76**, 091301 (2005).

[3] E. Seres, J. Seres, and C. Spielmann, *X-ray absorption spectroscopy in the keV range with laser generated high harmonic radiation*, Appl. Phys. Lett. **89**, 181919 (2006).

[4] M. E. Siemens, Q. Li, R. Yang, K. A. Nelson, E. H. Anderson, M. M. Murnane, and H. C. Kapteyn, *Quasi-ballistic thermal transport from nanoscale interfaces observed using ultrafast coherent soft X-ray beams*, Nat. Mater. **9**, 26 (2010).

[5] E. Turgut et al., *Controlling the Competition between Optically Induced Ultrafast Spin-Flip Scattering and Spin Transport in Magnetic Multilayers*, Phys. Rev. Lett. **110**, 197201 (2013).

[6] C. Ott, A. Kaldun, P. Raith, K. Meyer, M. Laux, J. Evers, C.H. Keitel, C. H. Greene, T. Pfeifer, Lorentz Meets Fano in Spectral Line Shapes: A Universal Phase and Its Laser Control, Science **340**, 716 (2013).

[7] J. Uhlig, et al., *Table-Top Ultrafast X-Ray Microcalorimeter Spectrometry for Molecular Structure*, Phys. Rev. Lett. **110**, 138302 (2013).

[8] W. Zhang et al., *Tracking excited-state charge and spin dynamics in iron coordination complexes*, Nature **509**, 345 (2014).

[9] J. Kern et al., *Taking snapshots of photosynthetic water oxidation using femtosecond X-ray diffraction and spectroscopy*, Nat. Commun. **5**, 4371 (2014).

[10] S. E. Canton et al., *Visualizing the non-equilibrium dynamics of photoinduced intramolecular electron transfer with femtosecond X-ray pulses*, Nat. Commun. **6**, 6359 (2015).

[11] R. Alonso-Mori, D. Sokaras, D. Zhu, T. Kroll, M. Chollet, Y. Feng, J. M. Glownia, J. Kern, H. T. Lemke, D. Nordlund, A. Robert, M. Sikorski, S. Song, T.-C. Weng, and U. Bergmann, *Photon-in photon-out hard X-ray spectroscopy at the Linac Coherent Light Source*, J. Synchrotron Radiat. **22**, 612 (2015).

[12] Z. Tao, C. Chen, T. Szilvási, M. Keller, M. Mavrikakis, H. Kapteyn, and M. Murnane, *Direct time-domain observation of attosecond final-state lifetimes in photoemission from solids*, Science **353**, 62 (2016).

[13] M. W. M. Jones, K. D. Elgass, M. D. Junker, M. D. de Jonge, G. A. van Riessen, *Molar concentration from sequential 2-D water-window X-ray ptychography and X-ray fluorescence in hydrated cells*, Sci. Rep. **6**, 24280 (2016).

[14] Luis Miaja-Avila et al., *Ultrafast Time-Resolved Hard X-Ray Emission Spectroscopy on a Tabletop*, Phys. Rev. X **6**, 031047 (2016).

[15] M. Chergui, E. Collet, *Photoinduced Structural Dynamics of Molecular Systems Mapped by Time-Resolved X-ray Methods*, Chem. Rev. **117**, 11025 (2017).

[16] G.C. O'Neil et al., *Ultrafast Time-Resolved X-ray Absorption Spectroscopy of Ferrioxalate Photolysis with a Laser Plasma X-ray Source and Microcalorimeter Array*, J. Phys. Chem. Lett. **8**, 1099 (2017).

[17] A. R. Attar, A. Bhattacherjee, C. D. Pemmaraju, K. Schnorr, K. D. Closser, D. Prendergast, S. R. Leone, *Femtosecond x-ray spectroscopy of an electrocyclic ring-opening reaction*, Science **356**, 54 (2017).

[18] Y. Pertot, C. Schmidt, M. Matthews, A. Chauvet, M. Huppert, V. Svoboda, A. von Conta, A. Tehlar, D. Baykusheva, J.-P. Wolf, H. J. Wörner, *Time-resolved x-ray absorption spectroscopy with a water window high-harmonic source*, Science **355**, 264 (2017).

[19] D. Popmintchev et al., *Near- and Extended-Edge X-Ray-Absorption Fine-Structure Spectroscopy Using Ultrafast Coherent High-Order Harmonic Supercontinua*, Phys. Rev. Lett. **120**, 093002 (2018).

[20] B. Buades, D. Moonshiram, Th. P. H. Sidiropoulos, I. León, P. Schmidt, I. Pi, N. Di Palo, S. L. Cousin, A. Picón, F. Koppens, J. Biegert, *Dispersive soft x-ray absorption fine-structure spectroscopy in graphite with an attosecond pulse*, Optica **5**, 502 (2018).

[21] P.B. Corkum, F. Krausz, Attosecond science, Nat. Phys. **3**, 381 (2007).

[22] F. Krausz, M. Ivanov, Attosecond physics, Rev. Mod. Phys. **81**, 163 (2009).

[23] F. Calegari, G. Sansone, S. Stagira, C. Vozzi, M. Nisoli, Advances in attosecond science, J. Phys. B: At. Mol. Opt. Phys. **49**, 062001 (2016).

[24] M.-C. Chen, P. Arpin, T. Popmintchev, M. Gerrity, B. Zhang, M. Seaberg, D. Popmintchev, M. M. Murnane, and H. C. Kapteyn, *Bright, Coherent, Ultrafast Soft X-Ray Harmonics Spanning the Water Window from a Tabletop Light Source*, Phys. Rev. Lett. **105**, 173901 (2010).

[25] T. Popmintchev, M. C. Chen, D. Popmintchev, P. Arpin, S. Brown, S. Ališauskas, G. Andriukaitis, T. Balčiunas, O. D. Mücke, A. Pugzlys et al., *Bright Coherent Ultrahigh Harmonics in the keV X-ray Regime from Mid-Infrared Femtosecond Lasers*, Science **336**, 1287 (2012).

[26] D. Popmintchev, C. Hernández-García, F. Dollar, C. Mancuso, J. A. Pérez-Hernández, M. C. Chen, A. Hankla, X. Gao, B. Shim, A. L. Gaeta et al., *Ultraviolet surprise: Efficient soft x-ray high-harmonic generation in multiply ionized plasmas*, Science **350**, 1225 (2015).

[27] J. Li, X. Ren, Y. Yin, Y. Cheng, E. Cunningham, Y. Wu, and Z. Chang, *Polarization gating of high harmonic generation in the water window*, Appl. Phys. Lett. **108**, 231102 (2016).

[28] J. Li, X. Ren, Y. Yin, K. Zhao, A. Chew, Y. Cheng, E. Cunningham, Y. Wang, Sh. Hu, Y. Wu, M. Chini, Z. Chang, *53-attosecond X-ray pulses reach the carbon K-edge*, Nat. Comm. **8**, 186 (2017).

[29] S.E. Harris, A.V. Sokolov, Subfemtosecond Pulse Generation by Molecular Modulation, Phys. Rev. Lett. **81**, 2894 (1998).

[30] M. Y. Shverdin, D. R. Walker, D. D. Yavuz, G. Y. Yin, and S. E. Harris, Generation of a Single-Cycle Optical Pulse, Phys. Rev. Lett. **94**, 033904 (2005).

[31] P. Tzallas, E. Skantzakis, L. A. A. Nikolopoulos, G. D. Tsakiris and D. Charalambidis, Extreme-ultraviolet pump–probe studies of one-femtosecond-scale electron dynamics, Nat. Phys. **7**, 781 (2011).

[32] E.J. Takahashi, P.i Lan, O.D. Mücke, Y. Nabekawa, K. Midorikawa, Attosecond nonlinear optics using gigawatt-scale isolated attosecond pulses, Nat. Commun. **4**, 2691 (2013).

[33] D. V. Korobkin, C. H. Nam, S. Suckewer, and A. Goltsov, Demonstration of Soft X-Ray Lasing to Ground State in Li III, Phys. Rev. Lett. **77**, 5206 (1996).

[34] J. J. Rocca, *Table-top soft x-ray lasers*, Rev. Sci. Instrum. **70**, 3799 (1999).

[35] H. Daido, *Review of soft x-ray laser researches and developments*, Rep. Prog. Phys. **65**, 1513 (2002).

[36] Ph. Zeitoun et al., *A high-intensity highly coherent soft X-ray femtosecond laser seeded by a high harmonic beam*, Nature **431**, 426 (2004).

[37] Y. Wang, E. Granados, M. A. Larotonda, M. Berrill, B. M. Luther, D. Patel, C. S. Menoni, and J. J. Rocca, *High-brightness injection-seeded soft-X-ray-laser amplifier using a solid target*, Phys. Rev. Lett. **97**, 123901 (2006).

[38] Y. Wang, E. Granados, F. Pedaci, D. Alessi, B. Luther, M. Berrill & J. J. Rocca, *Phase-coherent, injection-seeded, table-top soft-X-ray lasers at 18.9 nm and 13.9 nm*, Nat. Photonics **2**, 94 (2008).

[39] S. Suckewer and P. Jaegle, *X-Ray laser: past, present, and future*, Laser Phys. Lett. **6**, 411 (2009).

[40] E. Oliva, M. Fajardo, L. Li, S. Sebban, D. Ros, and Ph. Zeitoun, *Soft x-ray plasma-based seeded multistage amplification chain*, Opt. Lett. **37**, 4341 (2012).

[41] M. Nishikino and T. Kawachi, *X-ray laser plasma amplifiers*, Nat. Photonics **8**, 352 (2014).

[42] Y. Wang, S. Wang, E. Oliva, L. Li, M. Berrill, L. Yin, J. Nejdl, B. M. Luther, C. Proux, T. T. T. Le, J. Dunn, D. Ros, Ph. Zeitoun, and J. J. Rocca, *Gain dynamics in a soft-X-ray laser amplifier perturbed by a strong injected X-ray field*, Nat. Photonics **8**, 381 (2014).

[43] T. R. Akhmedzhanov, V. A. Antonov, A. Morozov, A. Goltsov, M. Scully, S. Suckewer, and O. Kocharovskaya, *Formation and amplification of subfemtosecond x-ray pulses in a plasma medium of hydrogenlike ions with a modulated resonant transition*, Phys. Rev. A **96**, 033825 (2017).

[44] Theoretical studies of hydrogen Rydberg atoms in electric fields, R. J. Damburg and V. V. Kolosov, in Rydberg States of Atoms and Molecules, edited by R. F. Stebbings , F. B. Dunning, Cambridge, UK: Cambridge University Press, 2011.
[45] T. R. Akhmedzhanov, M. Yu. Emelin, V. A. Antonov, Y. V. Radeonychev, M. Yu. Ryabikin, and O. Kocharovskaya, *Ultimate capabilities for few-cycle pulse formation via resonant interaction of XUV radiation with IR-field-dressed atoms*, Phys. Rev. A **95**, 023845 (2017).
[46] Y. Avitzour, S. Suckewer, *Feasibility of achieving gain in transition to the ground state of C VI at 3.4 nm*, J. Opt. Soc. Am. B **24**, 819 (2007).
[47] M. Gross and S. Haroche, Superradiance: an essay on the theory of collective spontaneous emission, Phys. Reports (Review Section of Physics Letters) **93**, 301 (1982).

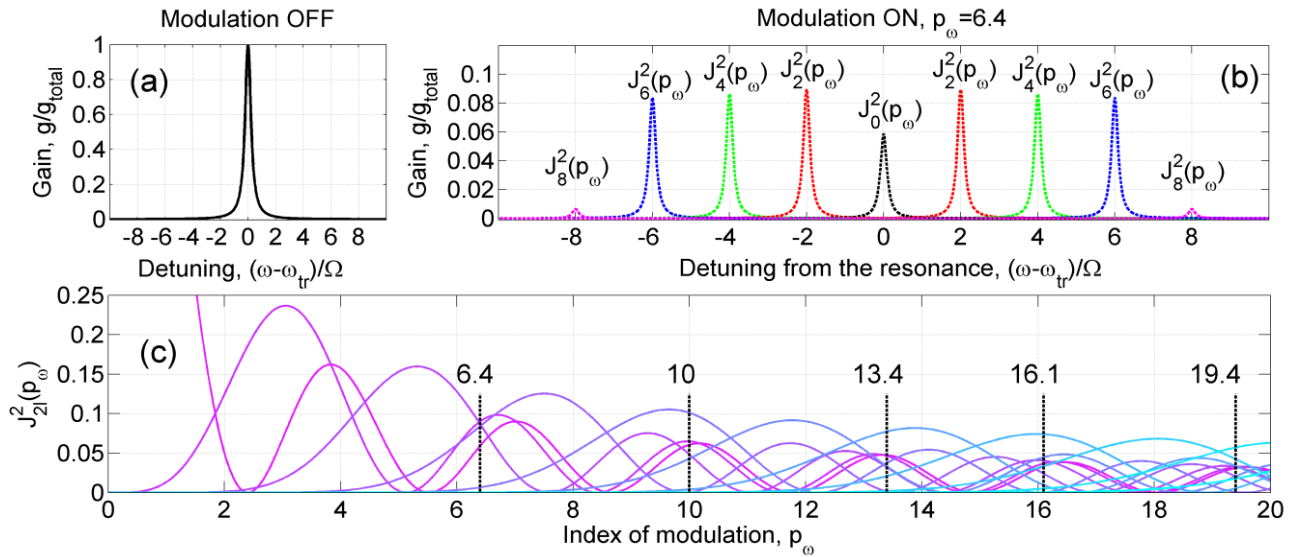

FIG. 1. (Color online) Frequency dependence of the gain for the incident X-ray field with (b) and without (a) modulating field. In panel (b) the value of modulation index is $p_\omega$=6.4; black, red, green, blue, and magenta curves correspond to the gain coefficients for "0-th", "±2-nd", "±4-ht", "±6-th", and "±8-th" sidebands of the inverted transition. Panel (c) shows dependence of the squared Bessel functions of even order (from "0-th" to "20-th" order) on the value of modulation index, $p_\omega$. The color reflects the order of Bessel function and is gradually changed from magenta, corresponding to $J_0^2(p_\omega)$, to cyan, which corresponds to $J_{20}^2(p_\omega)$. Vertical dashed lines indicate the values of $p_\omega$, at which the squares of Bessel functions of different orders are comparable to each other (compare with panel (b)).

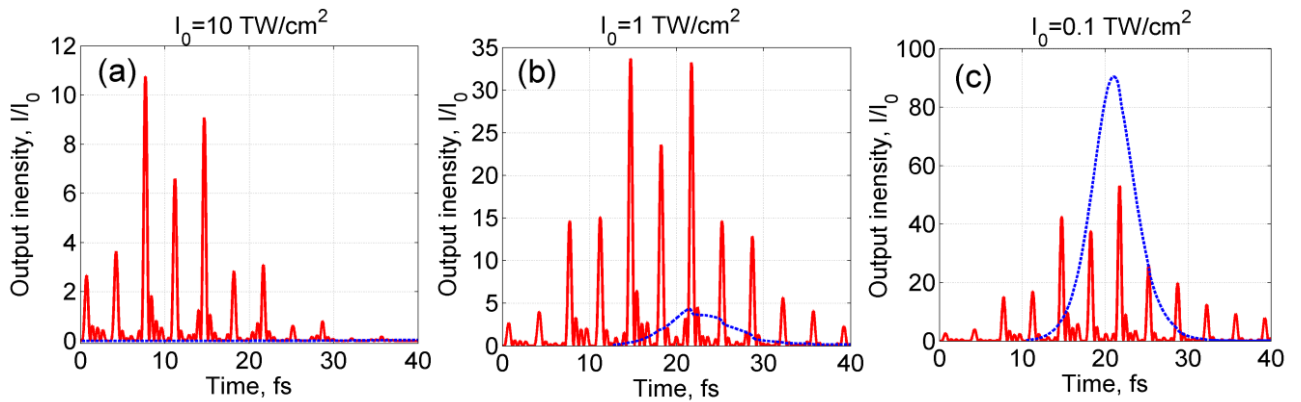

FIG. 2. (Color online) Time dependence of intensity of an amplified attosecond pulse train of $z$-polarization (red solid curve) and of an ASE of $y$-polarization (blue dashed curve) at the output from an active medium of $C^{5+}$ hydrogen-like X-ray laser. The length of active medium is $L=1$ mm, the concentration of $C^{5+}$ ions is $N_{ion}=10^{19}$ $1/cm^3$. Panels (a), (b), and (c) correspond to peak intensities of the incident attosecond pulse train $I_0=10^{13}$ W/cm$^2$, $10^{12}$ W/cm$^2$, and $10^{11}$ W/cm$^2$, respectively.

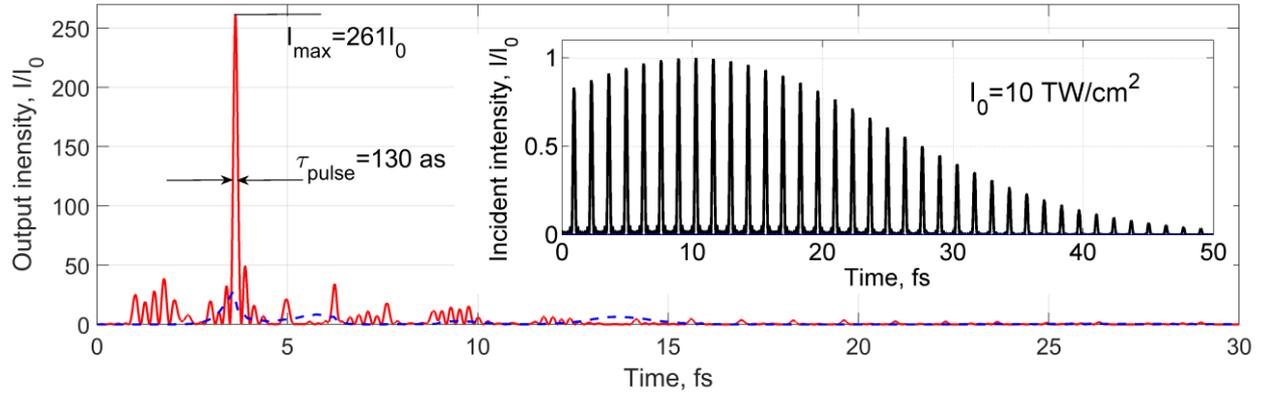

FIG. 3. (Color online) A single attosecond pulse in the "water window" spectral range, which is produced as a result of amplification of attosecond pulse train, shown in the inset, in an active medium of $C^{5+}$ hydrogen-like X-ray laser. Red solid curve is the amplified $z$-polarized X-ray signal; blue dashed curve corresponds to the ASE of $y$-polarization. The length of active medium is $L=7$ mm, the concentration of $C^{5+}$ ions is $N_{ion}=10^{19}$ $1/cm^3$.

# ATTOSECOND PULSE AMPLIFICATION IN A PLASMA-BASED X-RAY LASER DRESSED BY AN INFRARED LASER FIELD: SUPPLEMENTAL MATERIAL

V.A. Antonov[1,2,*], K.Ch. Han[3], T.R. Akhmedzhanov[3], Marlan Scully[3] and Olga Kocharovskaya[3]

[1]Institute of Applied Physics of the Russian Academy of Sciences, Nizhny Novgorod, Russia

[2]Prokhorov General Physics Institute of the Russian Academy of Sciences, Moscow, Russia

[3]Department of Physics and Astronomy, Texas A&M University, College Station, USA

**Abstract:** In this supplemental material we derive an analytical solution describing an amplification of the high-harmonic field in an active medium of a plasma-based X-ray laser in the linear regime within the three-level medium model and constant population inversion approximation. We provide also more detailed numerical study of an amplification process in both linear and nonlinear regimes taking into account a saturation effect within the five-level medium model.

## 1. Analytical solution describing the linear regime of amplification

In this section, we consider a transformation of the electric field of the harmonic of order $2(k+l)+1$ in the linear amplification regime (implying that the amplified field is relatively weak, so that it does not saturate the resonant transition and the population difference at this transition remains constant).

At the entrance to the medium, an electric field can be represented as

$$\vec{E}_{2(k+l)+1}(x \le 0, \tau) = \frac{1}{2}\vec{z}_0 E_{inc}^{[2(k+l)+1]} \exp\left\{-i\left(\bar{\omega}_{tr} + 2l\Omega\right)\tau\right\} + \text{c.c.}, \qquad (S1)$$

where $\vec{z}_0$ is a unit polarization vector along $z$-axis, $E_{inc}^{[2(k+l)+1]}$ is a complex amplitude of harmonic, $\tau = t - x/c$ is a local time in a reference frame propagating at the speed of light in vacuum, $c$, and c.c. stands for complex conjugation. During its propagation through the medium an electric field of "$2l$-th" harmonic acquires a form

$$\vec{E}_{2(k+l)+1}(x > 0, \tau) = \frac{1}{2}\vec{z}_0 \tilde{E}_{2(k+l)+1}(x, \tau) \exp\left\{-i\left(\bar{\omega}_{tr} + 2l\Omega\right)\tau\right\} + \text{c.c.}, \qquad (S2)$$

where its slowly-varying amplitude $\tilde{E}_{2(k+l)+1}(x, \tau)$ is given by

$$\tilde{E}_{2(k+l)+1}(x, \tau) = E_{inc}^{[2(k+l)+1]} \sum_{n=-\infty}^{\infty} e_{2n}^{(l)}(x) \exp\left\{-i2n\Omega\tau\right\}. \qquad (S3)$$

If polarization vectors of the harmonics and the modulating laser field coincide, then within the approximation of constant population difference at the transition $n=1 \leftrightarrow n=2$, propagation of the harmonic field through the medium can be described by the following equations for the three-level model taking into account the states $|1\rangle = |1s\rangle$, $|2\rangle = (|2s\rangle + |2p,m=0\rangle)/\sqrt{2}$, and $|3\rangle = (|2s\rangle - |2p,m=0\rangle)/\sqrt{2}$ [42]:

$$\begin{cases} \dfrac{\partial \tilde{E}_{2(k+l)+1}}{\partial z} = i\dfrac{2\pi\omega_k}{c}\tilde{P}_{2(k+l)+1}, \\ \tilde{P}_{2(k+l)+1} = 2N_{ion}d_{tr}\left[a_{21}^{(l)} - a_{31}^{(l)}\right], \\ \dfrac{\partial a_{21}^{(l)}}{\partial \tau} + \left(i\left[\bar{\omega}_{tr} - \omega_{2(k+l)+1} - \Delta_\omega \cos\left\{\Omega\left(\tau + \kappa\dfrac{z}{c}\right)\right\}\right] + \bar{\gamma}_{tr}\right)a_{21}^{(l)} = -\dfrac{in_{tr}d_{tr}}{2\hbar}\tilde{E}_{2(k+l)+1}, \\ \dfrac{\partial a_{31}^{(l)}}{\partial \tau} + \left(i\left[\bar{\omega}_{tr} - \omega_{2(k+l)+1} + \Delta_\omega \cos\left\{\Omega\left(\tau + \kappa\dfrac{z}{c}\right)\right\}\right] + \bar{\gamma}_{tr}\right)a_{31}^{(l)} = \dfrac{in_{tr}d_{tr}}{2\hbar}\tilde{E}_{2(k+l)+1}, \end{cases} \quad (S4)$$

where $\tilde{P}_{2(k+l)+1}$ is slowly-varying amplitude of the resonant polarization of the medium induced by the field of "2$l$-th" harmonic; $N_{ion}$ is density of the resonant ions; $d_{tr}$ is the absolute value of dipole moment of transitions $|1\rangle\leftrightarrow|2\rangle$ and $|1\rangle\leftrightarrow|3\rangle$; $a_{21}^{(l)}$ and $a_{31}^{(l)}$ are slowly-varying amplitudes of quantum coherencies $\rho_{21}$ and $\rho_{31}$, induced by the field of "2$l$-th" harmonic; $\Delta_\omega$ is amplitude of linear Stark shift induced by the modulating optical field; $\bar{\gamma}_{tr}$ is decoherence rate; $n_{tr} = \rho_{22} - \rho_{11} = \rho_{33} - \rho_{11}$ is population difference at the resonant transitions; and $\hbar$ is Planck's constant. Eqs. (S4) take into account plasma dispersion via the parameter $\kappa = \dfrac{\omega_{pl}^2}{2\Omega^2}$ (where $\omega_{pl}$ is plasma frequency for the modulating field), which characterizes the difference between the phase velocity of the laser field in the active medium and in vacuum.

Let us assume that for each harmonic order $2(k+l)+1$ the incident spectral component of the field (S2), (S3), dominates the sidebands (as it is shown below, this assumption is valid in the case of sufficiently strong plasma dispersion):

$$\left|e_0^{(l)}(x)\right| \gg \left|e_{2n}^{(l)}(x)\right| \quad (\forall l, n, x), \quad (S5)$$

Let us further consider the case

$$\bar{\gamma}_{tr} \ll \Omega. \quad (S6)$$

The value $\bar{\gamma}_{tr}$ is determined by the sum of collisional decoherence rate, $\gamma_{coll}$, ionization rate from the excited states $|2\rangle$ and $|3\rangle$ induced by modulating laser field, $w_{ion}^{(2,3)}$, and radiative decay rate, $\Gamma_{rad}$, namely, $\bar{\gamma}_{tr} = \gamma_{coll} + \left(w_{ion}^{(2,3)} + \Gamma_{rad}\right)/2$. Typically, $\gamma_{coll} \gg \Gamma_{rad}, w_{ion}^{(2,3)}$, while $w_{ion}^{(2,3)} < \Gamma_{rad}$. Further, $\Delta_\omega = 3\dfrac{m_e e^4}{\hbar^3 Z}\dfrac{E_C}{E_A}$ is the amplitude of linear Stark effect, $E_C$ is the strength of the modulating field, $E_A = m_e^2 e^5/\hbar^4 \cong 5.14\cdot 10^9 V/cm$ is the atomic unit of electric field strength, $e$ and $m_e$ are charge and mass of electron, respectively, $Z$ is nucleus charge number of the ions. Under the above assumptions, solution of Eqs. (S4) for boundary condition $\tilde{E}_{2(k+l)+1}(x=0,\tau) = E_{inc}^{[2(k+l)+1]}$ takes a form

$$e_0^{(l)}(x) = \exp\{g_{2l}x\}, \quad (S7)$$

$$e_{2n}^{(l)}(x) = \dfrac{J_{2(n+l)}(p_\omega)}{J_{2l}(p_\omega)}\dfrac{\exp\{g_{2l}[1-i2n\alpha_l]x\}-1}{1-i2n\alpha_l}, \text{ if } n \neq 0. \quad (S8)$$

$$g_{2l} = g_{total}J_{2l}^2(p_\omega) \quad (S9)$$

In these equations, $g_{total} = \frac{4\pi n_{tr} N_{ion} d_{tr}^2 \bar{\omega}_{tr}}{\hbar \bar{\gamma}_{tr} c}$ is an increment of amplification of the resonant X-ray field at the frequency $\bar{\omega}_{tr}$ without linear Stark effect, $J_m(x)$ is Bessel function of the first kind of order $m$, $p_\omega = \Delta_\omega / \Omega$ is the modulation index, and $\alpha_l = \frac{\kappa}{g_{total} J_{2l}^2(p_\omega)} \frac{\Omega}{c}$ is a dimensionless parameter, which can be represented as $\alpha_l = \pi \frac{L_{amp}^{(l)}}{L_{coh}}$, where $L_{amp}^{(l)} = \frac{1}{g_{total} J_{2l}^2(p_\omega)}$ is the amplification length of "2l-th" harmonic and $L_{coh} = \lambda_C \frac{\Omega^2}{\omega_{pl}^2}$ is the coherence length, i.e. the length of coherent interaction between the high-harmonic field and the modulating optical field (the length at which a phase shift of π is acquired by an IR field (with respect to an X-ray field) due to the plasma dispersion in an active medium), $\lambda_C$ is the wavelength of the modulating field.

If the incident X-ray field consists of several high-order harmonics of orders ranging from $2(k-l_{min})+1$ to $2(k+l_{max})+1$,

$$\vec{E}_{X-ray}(x \leq 0, \tau) = \frac{1}{2} \vec{z}_0 \sum_{l=-l_{min}}^{l_{max}} E_{inc}^{[2(k+l)+1]} \exp\{-i\omega_{2(k+l)+1}\tau\} + c.c., \quad (S10)$$

then, within the approximation of fixed population differences at the transitions $|1\rangle \leftrightarrow |2\rangle$ and $|1\rangle \leftrightarrow |3\rangle$, each harmonic will propagate through the medium independently of the others, being amplified and generating sidebands at the frequencies of other harmonics in accordance with Eqs. (S2), (S3), and (S7)-(S9). In such a case, the X-ray field within the medium takes a form

$$\vec{E}_{X-ray}(x, \tau) = \frac{1}{2} \vec{z}_0 \sum_{l=-l_{min}}^{l_{max}} E_{inc}^{[2(k+l)+1]} \sum_{n=-\infty}^{\infty} e_{2n}^{(l)}(x) \exp\{-i(\bar{\omega}_{tr} + 2[l+n]\Omega)\tau\} + c.c. \quad (S11)$$

In order to amplify a large number of harmonics, one needs to have appreciably nonzero values of $g_{2l}$ for various $l$, see Eq. (S9). This requires a large value of the modulation index $p_\omega$, which can be achieved either via increasing $E_C$, or via reducing $\Omega$. As it follows from Eq. (S8), in the case of sufficiently strong plasma dispersion, which corresponds to $\alpha_l \gg 1$, rescattering of harmonics into each other is negligible, $|e_{2n}^{(l)}(x)| \ll |e_0^{(l)}(x)|$, which justifies Eq. (S5). In such a case

$$\vec{E}_{X-ray}(x, \tau) = \frac{1}{2} \vec{z}_0 \sum_{l=-l_{min}}^{l_{max}} E_{inc}^{[2(k+l)+1]} \exp\{g_{2l} x\} \exp\{-i(\bar{\omega}_{tr} + 2l\Omega)\tau\} + c.c. \quad (S12)$$

As follows from Eqs. (S9) and (S12), the amplification coefficient of "2l-th" harmonic is proportional to the squared Bessel function of order $2l$ of modulation index $p_\omega$, $J_{2l}^2(p_\omega)$; each harmonic of the incident X-ray field is amplified independently of the others, and relative phases of harmonics remain constant during their propagation through the medium. In particular, if at the entrance to the medium the harmonics (S10) are phase-aligned and correspond to a train of attosecond pulses, these pulses will be amplified as a whole.

A number of harmonics which can be simultaneously amplified increases with increasing value of the modulation index $p_\omega$ (since for a larger argument more Bessel functions have appreciably nonzero values). Particularly, the values $p_\omega = 6.4$, $p_\omega = 10$, $p_\omega = 13.4$, $p_\omega = 16.1$, and $p_\omega = 19.4$ are suitable for amplification of harmonics with even numbers ranging from "-6-th" to "6-th", from "-10-th" to "10-th", from "-14-th" to "14-th", from "-16-th" to "16-th", and from "-20-th" to "20-th", respectively. Nevertheless, with increasing value of the modulation index, an average of the gain coefficient, $g_{total} J_{2l}^2(p_\omega)$, over harmonic number $2l$ decreases as $1/p_\omega$, while the differences between amplification coefficients corresponding to different numbers $2l$ grow, as it shown in Fig. S1, where we plot $J_{2l}^2(p_\omega)$ for $p_\omega = 6.4$, $p_\omega = 10$, and $p_\omega = 19.4$. Thus, with increasing number of amplified harmonics the amplification becomes less efficient, while distortions of spectrum (and of time dependence) of the amplified X-ray signal grow.

## 2. Numerical analysis taking into account the nonlinear effects

In order to verify the predictions of the analytical theory and to describe the nonlinear stage of amplification, we perform a numerical study of a more comprehensive model, which takes into account a finite duration of the incident X-ray field, variation of population differences between the resonant atomic states $|1\rangle$, $|2\rangle$, and $|3\rangle$, as well as influence of the states $|4\rangle = |2p, m = 1\rangle$ and $|5\rangle = |2p, m = -1\rangle$, which are coupled to the ground state $|1\rangle$ by $y$-polarized X-ray field (spontaneously emerging in the active medium). This model includes a wave equation for the X-ray field consisting of two polarization components, $\vec{E}_{X-ray}(x,t) = \vec{z}_0 E_z(x,t) + \vec{y}_0 E_y(x,t)$:

$$\frac{\partial^2 \vec{E}_{X-ray}}{\partial x^2} - \frac{1}{c^2}\frac{\partial^2 \vec{E}_{X-ray}}{\partial t^2} = \frac{4\pi}{c^2}\frac{\partial^2 \vec{P}}{\partial t^2}, \tag{S13}$$

as well as a relation between the macroscopic polarization of the medium and the elements of density matrix of the resonant ions:

$$\vec{P}(x,t) = N_{ion}\left(\vec{d}_{12}\rho_{21} + \vec{d}_{13}\rho_{31} + \vec{d}_{14}\rho_{41} + \vec{d}_{15}\rho_{51}\right) + \text{c.c.}, \tag{S14}$$

and equations for the density-matrix elements:

$$\dot{\rho}_{11} = +\gamma_{11}(\rho_{22} + \rho_{33} + \rho_{44} + \rho_{55}) - i[H,\rho]_{11},$$
$$\dot{\rho}_{ij} = -\gamma_{ij}\rho_{ij} - i[H,\rho]_{ij}, \quad ij \neq 11,$$

$$H = \begin{pmatrix} E_1 & -E_z d_{tr} & E_z d_{tr} & -iE_y d_{tr} & -iE_y d_{tr} \\ -E_z d_{tr} & E_2 - \Delta_\omega \cos(\Omega[t - xn_{pl}/c]) & 0 & 0 & 0 \\ E_z d_{tr} & 0 & E_3 + \Delta_\omega \cos(\Omega[t - xn_{pl}/c]) & 0 & 0 \\ iE_y d_{tr} & 0 & 0 & E_4 & 0 \\ iE_y d_{tr} & 0 & 0 & 0 & E_5 \end{pmatrix}, \tag{S15}$$

where H is Hamiltonian of the ions, $E_i$ is energy of the "$i$-th" state (which takes into account quadratic Stark shift induced by the modulating optical field), $n_{pl} = \sqrt{1 - \omega_{pl}^2/\Omega^2}$ is plasma refraction index for the modulating field, $\omega_{pl}^2 = \frac{4\pi N_e e^2}{m_e}$, $N_e$ is density of free electrons, and $\gamma_{ij}$ is relaxation rate of the density matrix element $\rho_{ij}$. As initial conditions, we assume that at $\tau = 0$ all the ions are excited to the states $|2\rangle$-$|5\rangle$ with equal probability (thus implying excitation by running wave of a pump). Spontaneous emission at the transitions $|i\rangle \to |1\rangle$, $i=2,3,4,5$, is taken into account via introduction of random initial values of quantum coherencies at these transitions, $\rho_{i1}^{(0)} \equiv \rho_{i1}(\tau = 0)$, following Gross-Haroche approach [47]. Specifically, the medium is split into a set of thin slices of the length $l \leq l_{crit}$, where $l_{crit} = \sqrt{\frac{8\pi c}{3\lambda_{tr}^2 \Gamma_{rad}^{(ii)} N_{ion} \rho_{ii}^{(0)}}}$, $\Gamma_{rad}^{(ii)}$ is radiative lifetime of the state $|i\rangle$, $\rho_{ii}^{(0)} \equiv \rho_{ii}(\tau = 0)$ is initial population of this state, and $\lambda_{tr} = 2\pi c/\bar{\omega}_{tr}$ is wavelength of the resonant X-ray field. The initial values of the quantum coherencies are set independently in each slice as $\rho_{i1,m}^{(0)} = \frac{C_i}{2N\pi R^2 l} A_{i1m} e^{j\varphi_{i1m}}$, where $m$ is a number of slice centered at

$x_m = (m - 1/2)l$, $C_{2,3} = \pm 1$ and $C_{4,5} = -j$ are the coefficients, which account for the difference in dipole moments of transitions $|i\rangle \rightarrow |1\rangle$ for different values of $i$; while the amplitude $0 \leq A_{i1m} < \infty$ and the phase $0 \leq \varphi_{i1m} < 2\pi$ are random values. The constants $A_{i1m}$ and $\varphi_{i1m}$ satisfy the following probability distributions: $p(A_{i1m}^2) = \frac{1}{N_{il}} e^{-\frac{A_{i1m}^2}{N_{il}}}$, where $N_{il} = \rho_{ii}^{(0)} N_{ion} \pi R^2 l$ is a number of ions in the state $|i\rangle$ in the slice number $m$ and $R$ is radius of the plasma channel, and $p(\varphi_{i1m}) = 1/2\pi$. As an incident $z$-polarized X-ray field, we consider a train of attosecond pulses with Gaussian envelope centered at $t_{peak}$ with duration (the full width at half maximum of intensity) $t_{1/2}$:

$$E_z(t, x=0) = \frac{1}{2} E_{hh} \exp\left\{-2\ln 2(t - t_{peak})^2 / t_{1/2}^2\right\} \sum_{l=-l_{max}}^{l=l_{max}} \exp\left\{-i(\bar{\omega}_{tr} + 2l\Omega)t\right\} + \text{c.c.} \tag{S16}$$

Eq. (S16) is identical to Eq. (4) of the paper and implies that at the entrance to the medium all the spectral components of the seeding field are phase-aligned and have identical amplitudes.

As it was mentioned before, $z$-polarized seeding X-ray field is amplified due to transitions from the excited states $|2\rangle$ and $|3\rangle$, dressed by the modulating laser field, to the ground state $|1\rangle$. The amplification occurs at multiple frequencies $\omega_{2l} = \bar{\omega}_{tr} + 2l\Omega$, $l=0,\pm 1,\pm 2,...$, and the amplification coefficient $g_l = g_{total} J_{2l}^2(p_\omega)$ at each particular frequency, $\omega_{2l}$, is $1/J_{2n}^2(p_\omega)$ times smaller than the amplification coefficient for $z$-polarized signal without the modulating field, $g_{total}$. At the same time, the amplification for $y$-polarized ASE occurs due to transitions from the states $|4\rangle$ and $|5\rangle$, which correspond to degenerate energy levels $E_{4,5} = E_4 = E_5$, to the ground state $|1\rangle$ at the frequency $\bar{\omega}_{tr}^{(ASE)} = E_{4,5} - E_1$ (the value of $\bar{\omega}_{tr}^{(ASE)}$ differs from $\bar{\omega}_{tr}$ because of the difference in quadratic Stark shifts of the states $|2\rangle$, $|3\rangle$, and $|4\rangle$, $|5\rangle$). The amplification coefficient for the $y$-polarized ASE equals $g_{total}$ and thus is several times larger than the amplification coefficients for $z$-polarized harmonics. Therefore, in the linear amplification regime the $y$-polarized ASE grows faster than the $z$-polarized seeding X-ray field and may extract the major part of the energy, initially stored in the population inversion of the medium. In order to prevent it, the incident X-ray field (S16) should deplete population inversion at the transitions $|2\rangle \leftrightarrow |1\rangle$ and $|3\rangle \leftrightarrow |1\rangle$ thus increasing population of the ground state $|1\rangle$ and reducing (or even preventing) the gain at the transitions $|4\rangle \leftrightarrow |1\rangle$ and $|4\rangle \leftrightarrow |1\rangle$ before the $y$-polarized ASE becomes substantial. For this purpose the incident $z$-polarized X-ray field (i) should enter the medium prior to the development of the ASE and (ii) should be sufficiently strong.

The dependence of parameters of the amplified X-ray field on the delay of the seeding $z$-polarized attosecond pulse train (S16) with respect to the time zero (the time, at which the population inversion is created by a pump laser pulse), is illustrated by Figs. S2-S4. In all these figures we assume $I_0$=1 TW/cm$^2$ similarly to Fig. 2(b) of the paper. In Fig. S2 we plot the time dependence of intensity of the seeding X-ray field for $t_{peak}$ =0 fs, $t_{peak}$ =20 fs, and $t_{peak}$ =40 fs in panels (a), (b), and (c), respectively. The incident X-ray field in the case $t_{peak}$ =10 fs is shown on an inset to Fig. 3 of the paper (but for the different repetition period of attosecond pulses). The parameters of the medium are the same as in Fig. 2 of the paper, that is, density of the resonant ions is $N_{ion} = 10^{19} cm^{-3}$, density of free electrons is $N_{el} = 15 N_{ion}$, and the medium length is $L$=1 mm. In Fig. S3 we show the time dependences of populations of the relevant states of the ions, $\rho_{ii}$, $i$=1,2,...,5, at the centre of the medium, $x$=$L/2$=0.5 mm, which correspond to the same values of delay: $t_{peak}$=0 fs (a), $t_{peak}$ =20 fs (b), and $t_{peak}$ =40 fs (c) as in Fig. S2 (the case $t_{peak}$ =10 fs is illustrated by Fig. S5(b)). Finally, the time dependences of intensity of (i) the amplified $z$-polarized attosecond pulse train and (ii) the ASE of $y$-polarization at the output of the medium, $x$=$L$=1 mm, for different values of $t_{peak}$ are shown in Fig. S4. Panels (a), (b), and (c) correspond to $t_{peak}$ =0 fs, $t_{peak}$ =20 fs, and $t_{peak}$ =40 fs, respectively. The case $t_{peak}$ =10 fs is presented in panel (b) of Fig. 2 of the main text. As follows from these figures, the optimum value of delay of the incident X-ray field with respect to the moment of creation of population differences at the inverted transitions is $t_{peak}$ =10 fs. For the larger values of delay, the $y$-polarized ASE starts to play the role, while for smaller delays of the seeding X-ray field the amplification is less efficient, since the effective duration of the incident X-ray field (at positive times) becomes comparable to the response time of the resonant polarization of the active medium (which is~ $1/\gamma_{21} \approx$ 20 fs in the considered case). It is worth noting that, since $y$-polarized ASE has a stochastic nature, its properties (timing, peak intensity, and pulse shape) vary considerably (by several tens or hundreds of percent) from shot to shot, however, as long as the incident X-ray field is sufficiently strong and saturates the transitions $|1\rangle \leftrightarrow |2\rangle$ and $|1\rangle \leftrightarrow |3\rangle$ before the development of $y$-polarized ASE, this variation has very limited effect on $z$-polarized amplified signal.

In Fig. S5 we plot the time dependencies of populations of the relevant states, $|1\rangle$-$|5\rangle$, of the ions, for the same parameters of the medium and the incident X-ray field, as in Figs. S2-S4 and Fig. 2 of the paper. Panels (a), (b), and (c) correspond to the peak intensities of the incident attosecond pulse train $I_0$=0.1 TW/cm$^2$, $I_0$=1 TW/cm$^2$, and $I_0$=10 TW/cm$^2$, respectively. The time dependencies of populations are plotted for the center of the medium, $x=L/2=0.5$ mm, and correspond to the optimal value of $t_{peak}$ =10 fs. As follows from Fig. S5, with increasing intensity of the incident X-ray field, the energy transfer from the active medium into the amplified attosecond pulse train becomes more efficient. Particularly, it follows from the fact that the difference between the values $\rho_{22}$, $\rho_{33}$, and $\rho_{44}$, $\rho_{55}$ at $\tau \to \infty$ grows with increasing $I_0$. As a result, the absolute value of peak intensity of the amplified pulse train in the case $I_0$=10 TW/cm$^2$ is approximately 3.3 times higher than in the case $I_0$=1 TW/cm$^2$, and approximately 20 times higher than in the case $I_0$=1 TW/cm$^2$, see Fig. 2 of the paper.

Figs. S2-S5 and Fig. 2 of the paper were plotted for the value of modulation index $p_\omega$=6.4, which corresponds to the gain coefficients for the spectral components of $z$-polarized X-ray field (high-order harmonics of different orders), shown in Fig. S1 (a). In Fig. S6 we address the case of much higher modulation index, $p_\omega$=19.4, which provides a possibility to amplify larger number of harmonics at the cost of lower gain coefficients, see Fig. S1 (c). In Fig. S6 the thickness of the medium is the same as before, $L$=1 mm, but unlike Fig. 2 and Figs. S2-S5, the modulating 2.1 μm IR field has higher intensity ($I_C$=2.5×10$^{16}$ W/cm$^2$ instead of $I_C$=2.7×10$^{15}$ W/cm$^2$), while the incident X-ray field consists of 21 phase-aligned high-order harmonics of the modulating field (instead of 7 harmonics in Fig. 2 and Figs. S2-S5). The peak intensity of the seeding field is $I_0$=1 TW/cm$^2$ and the delay of the peak of its envelope from the time zero is $t_{peak}$ =10 fs. As follows from comparison of Fig. S6 with Fig. 2(b), the increase in modulation index allows to amplify shorter pulses but with lower gain, in full agreement with Fig. S1.

In Figs. S7-S9 we consider the possibility to isolate a single attosecond pulse in optically deep active medium of an X-ray laser. All the parameters are the same as in Fig. 3 of the paper: the wavelength and intensity of the modulating field are $\lambda_C$=801.53 nm and $I_C$=1.9×10$^{16}$ W/cm$^2$, respectively, the incident X-ray field consists of a set of 7 in-phase harmonics of the modulating field (of orders 231 to 243) with peak intensity $I_0$=10 TW/cm$^2$ under the Gaussian envelope with the duration $t_{1/2}$=35 fs and the delay of its peak from the time zero $t_{peak}$ =10 fs; the length of the active medium is $L$=7 mm. In Fig. S7 we show the evolution of time dependences of both the $z$-polarized amplified attosecond pulse train (a), and the $y$-polarized ASE (b) during their propagation through the medium. The time dependencies are plotted for the entrance to the medium, $x$=0, and for different propagation distances: $x$ = 1 mm, 2 mm,..., including the exit from the medium, $x$ = 7 mm. The results for $x$=1 mm are nearly the same as in Fig. 2 (c) of the paper. However, with increasing propagation distance the amplified attosecond pulse train is intensified, resulting in faster depletion of the population differences at the resonant transitions of the ions, $|2\rangle \leftrightarrow |1\rangle$ and $|3\rangle \leftrightarrow |1\rangle$, see Fig. S8 and Fig. S9. As a result, the gain of the active medium is temporally confined to a short time-interval near $\tau$=0, which is reduced with increasing propagation distance. Already at the center of the medium, $x$=3.5 mm, this interval becomes shorter than 5 fs, allowing for amplification of only a few pulses from the front edge of the incident attosecond pulse train. Yet, the active medium has a finite response time to the resonant X-ray field. Since the medium is optically deep, $g_{total}J_0^2(p_\omega)L = 22.7$, its response time is considerably shorter than the lifetimes of quantum coherencies at the resonant transitions, $1/\gamma_{21}= 1/\gamma_{31}\approx 20$ fs. Nevertheless, few femtoseconds are needed for the gain of the active medium to build-up. Therefore, till the population differences are nearly constant, each consequent attosecond pulse from the incident pulse train is amplified stronger than its predecessor. For this reason, at $x = 7$ mm the third attosecond pulse from the train is amplified much more efficiently than the first two pulses and depletes the population differences at the resonant transitions, preventing the amplification of the next pulses. As a result, a large part of the energy, initially kept in the population difference of the active medium, is transferred into a single attosecond pulse with duration 130 as, which is amplified up to peak intensity $I_{max} = 261 \times I_0 \approx 2.6 \times 10^{15}$ W/cm$^2$.

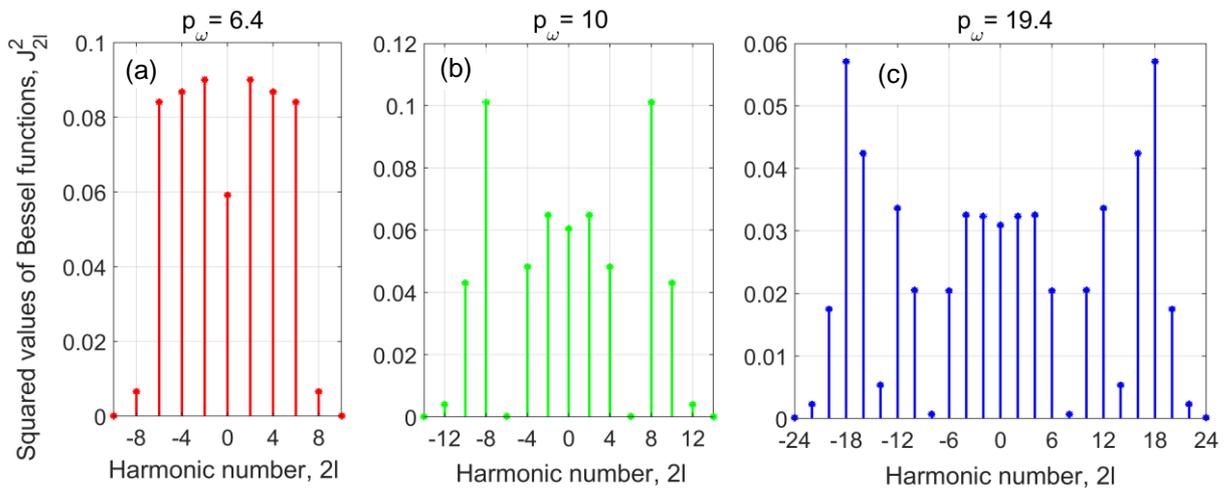

FIG. S1. The gain coefficients for the spectral components of the incident X-ray field (high-order harmonics of different orders) (S9) for the different values of modulation index: $p_\omega$=6.4 (a), $p_\omega$=10 (b), and $p_\omega$=19.4 (c). With increasing value of the modulation index the total number of amplified spectral components grows at the cost of (i) reduced magnitude of the gain coefficients and (ii) less uniform amplification.

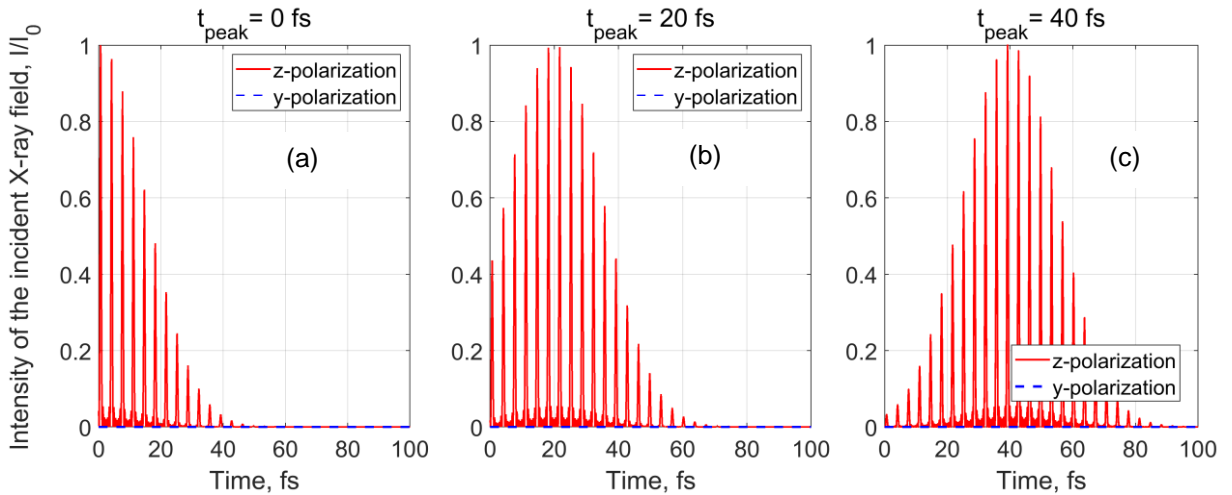

FIG. S2. Time dependence of intensity of the incident X-ray field (S16) for different values of delay of the peak of its envelope from the time zero: $t_{peak}$ =0 fs (a), 20 fs (b), and 40 fs (c). Red solid curve corresponds to z-polarization, blue dashed curve shows y-polarization (which is zero at the entrance to the medium). The intensity is normalized to its peak value.

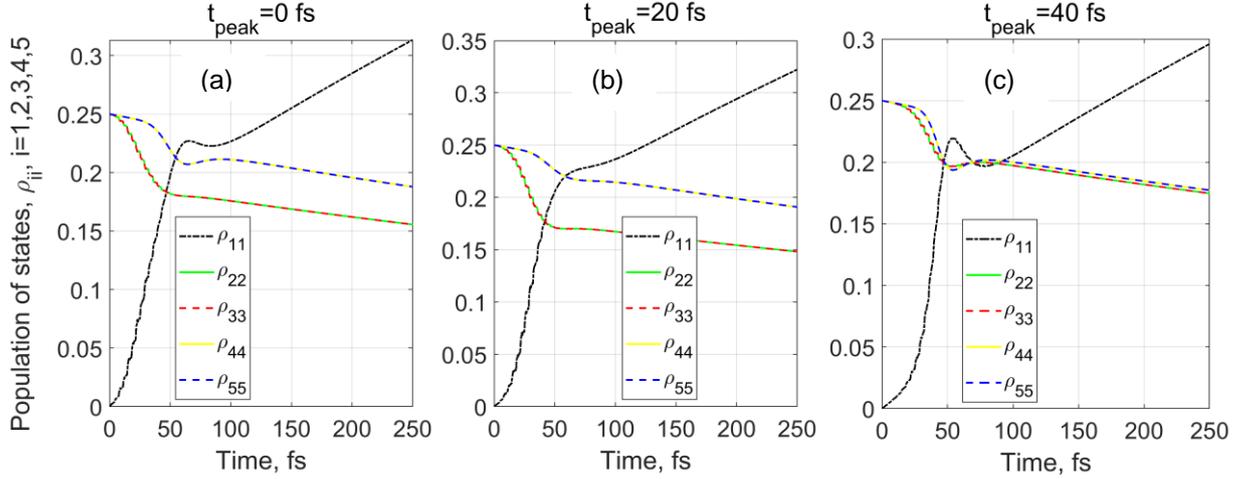

FIG. S3. Time dependencies of populations of the relevant states of the ions: $\rho_{11}$ (black dash-dotted curve), $\rho_{22}$ (green solid curve), $\rho_{33}$ (red dashed curve), $\rho_{44}$ (yellow solid curve), and $\rho_{55}$ (blue dashed curve) at x=0.5 mm. The peak intensity of the incident attosecond pulse train (S16) is $I_0$=1 TW/cm$^2$. Panels (a), (b) and (c) correspond to $t_{peak}$ =0 fs, 20 fs, and 40 fs, respectively. The time dependencies of $\rho_{22}$ and $\rho_{33}$, as well as of $\rho_{44}$ and $\rho_{55}$, in pairs almost coincide with each other.

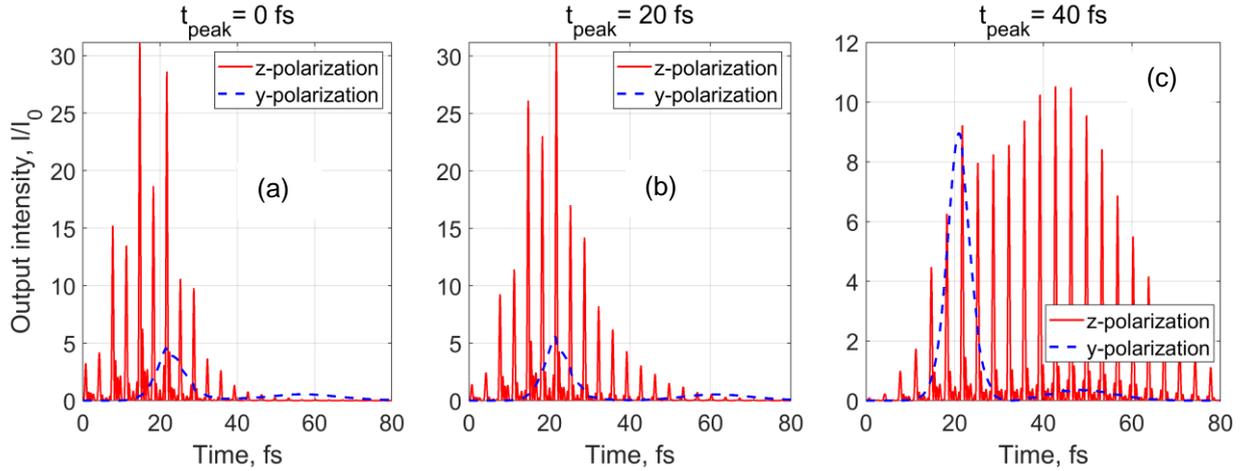

FIG. S4. Time dependence of intensity of the X-ray field at the output of 1 mm long active medium of C$^{5+}$ hydrogen-like X-ray laser. Red solid curve corresponds to the z-polarized amplified attosecond pulse train; blue dashed curve shows ASE of y-polarization. Panels (a), (b) and (c) correspond to different delays of the incident X-ray field (S16) with respect to the time zero: $t_{peak}$ =0 fs (a), 20 fs (b), and 40 fs (c). The intensity is normalized to the peak intensity of the incident X-ray field, $I_0$.

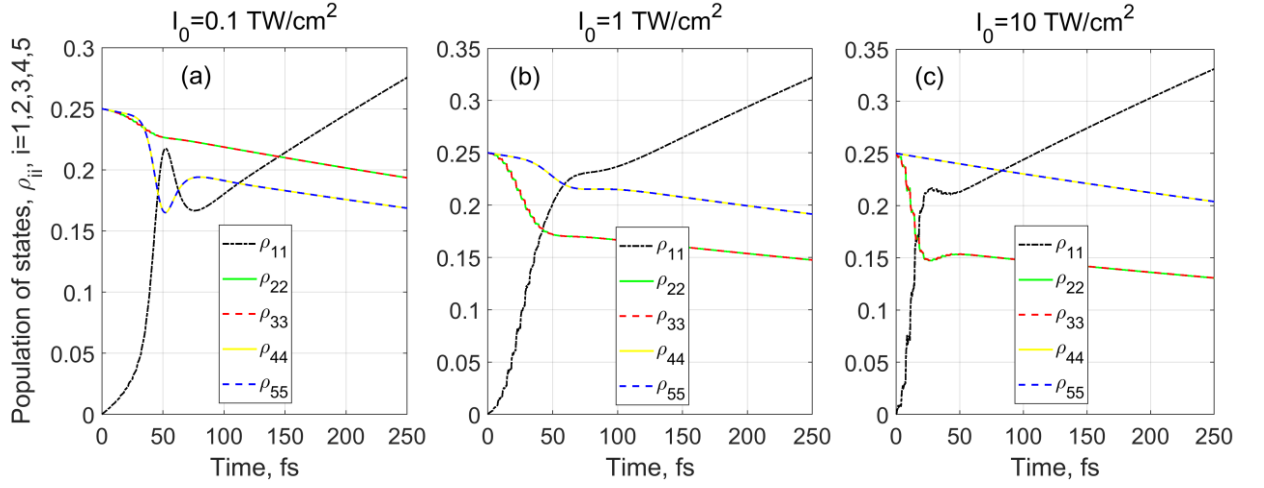

FIG. S5. Time dependencies of populations of the relevant states of the ions: $\rho_{11}$ (black dash-dotted curve), $\rho_{22}$ (green solid curve), $\rho_{33}$ (red dashed curve), $\rho_{44}$ (yellow solid curve), and $\rho_{55}$ (blue dashed curve) at $x$=0.5 mm. Delay of the incident X-ray field (S16) with respect to the time zero is $t_{peak}$ =10 fs. Panels (a), (b), and (c) correspond to the different peak intensities of the incident X-ray field: $I_0$=0.1 TW/cm$^2$ (a), 1 TW/cm$^2$ (b), and 10 TW/cm$^2$ (c). The time dependencies of $\rho_{22}$ and $\rho_{33}$, as well as of $\rho_{44}$ and $\rho_{55}$, in pairs almost coincide with each other.

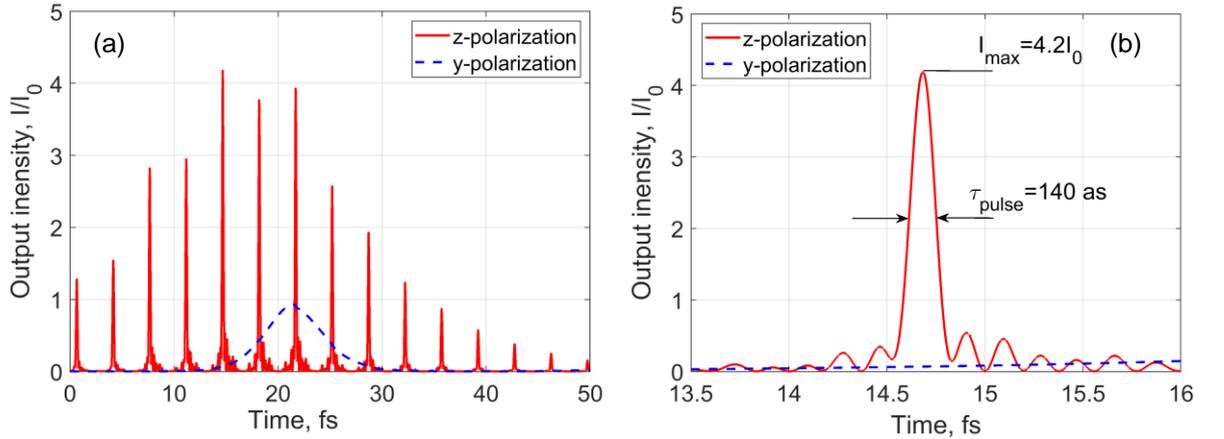

FIG. S6. Time dependence of intensity of the $z$-polarized amplified attosecond pulse train (red solid curve) and of the $y$-polarized ASE (blue dashed curve) at the output of an active medium of $C^{5+}$ hydrogen-like X-ray laser. The incident X-ray field consists of 21 in-phase harmonics of the modulating optical field with 2.1 µm wavelength. The parameters of the seeding pulse are $t_{peak}$ =10 fs, $I_0$=1 TW/cm$^2$. The intensity of the modulating field is $I_C$=2.5×10$^{16}$ W/cm$^2$. The length of active medium is $L$=1 mm, the density of $C^{5+}$ ions is $N_{ion}$=10$^{19}$ cm$^{-3}$. Panels (a) and (b) show the same time dependence with different time-scale. The intensity is normalized to the peak intensity of the incident X-ray field, $I_0$.

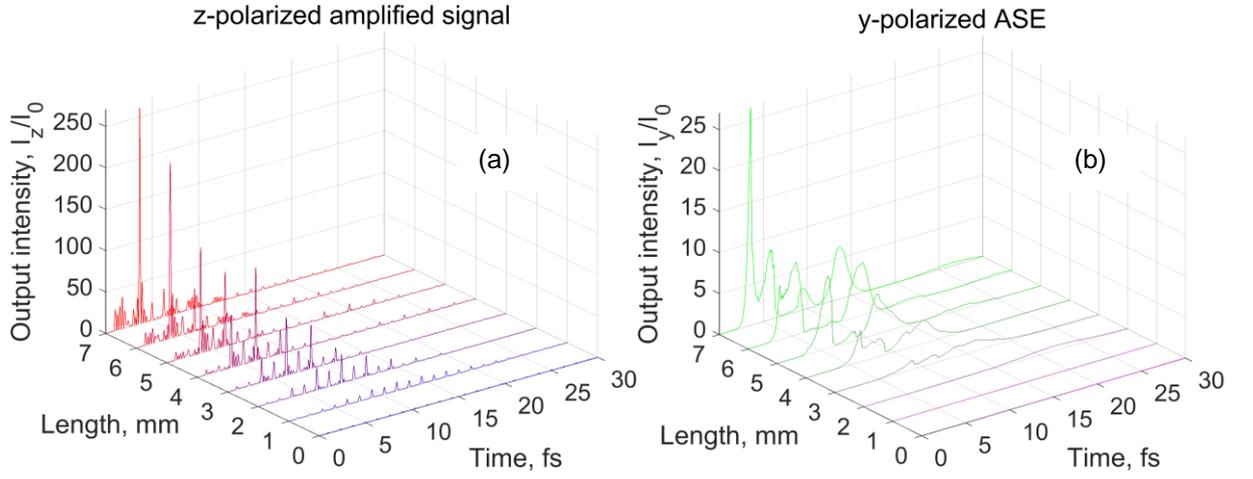

FIG. S7. Time dependencies of intensity of the amplified attosecond pulse train of $z$-polarization, panel (a), and of the ASE of $y$-polarization, panel (b), plotted for different lengths of the active medium of $C^{5+}$ hydrogen-like X-ray laser ($L$=1 mm, 2 mm, 3 mm, 4 mm, 5 mm, 6 mm, and 7 mm). The color is gradually changed from blue to red in panel (a) and from magenta to green in panel (b) for better visibility. The intensities of both polarization components are normalized to the peak intensity of the incident X-ray field, $I_0$.

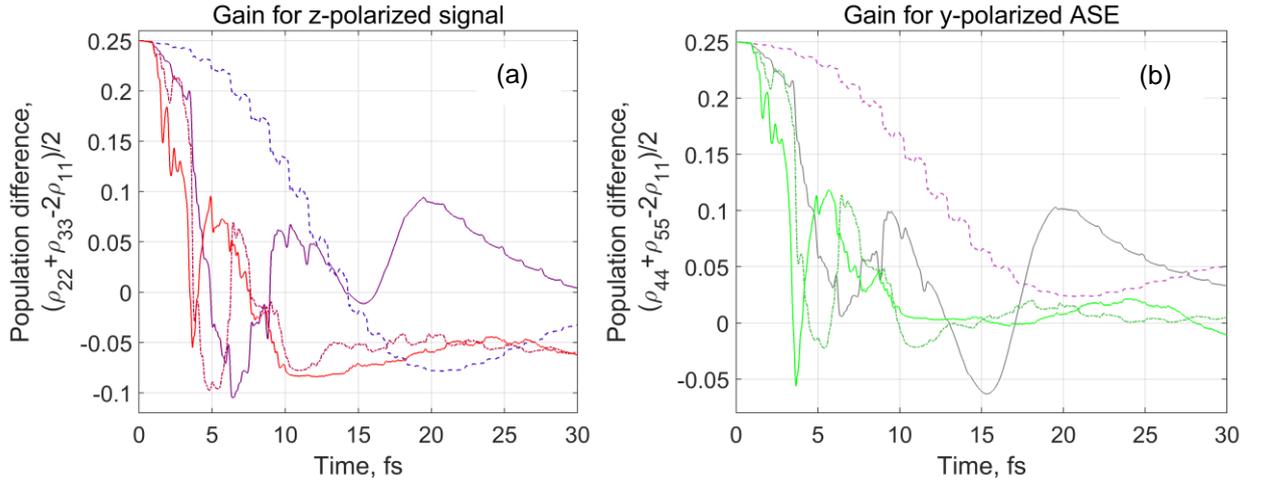

FIG. S8. Time dependencies of the population differences between the excited states $|2\rangle$, $|3\rangle$, and the ground state $|1\rangle$, panel (a), as well as the time dependencies the population differences between the excited states $|4\rangle$, $|5\rangle$, and the state $|1\rangle$, panel (b), plotted for different lengths ($L$=1 mm, 3 mm, 5 mm, and 7 mm) of the active medium of $C^{5+}$ hydrogen-like X-ray laser. The color is gradually changed from blue to red in panel (a) and from magenta to green in panel (b) for better visibility. The curves, which correspond to $L$=1 mm and $L$=5 mm, are plotted by dashed and dash-dotted lines, respectively. The curves corresponding to $L$=3 mm and $L$=7 mm are plotted by solid lines.

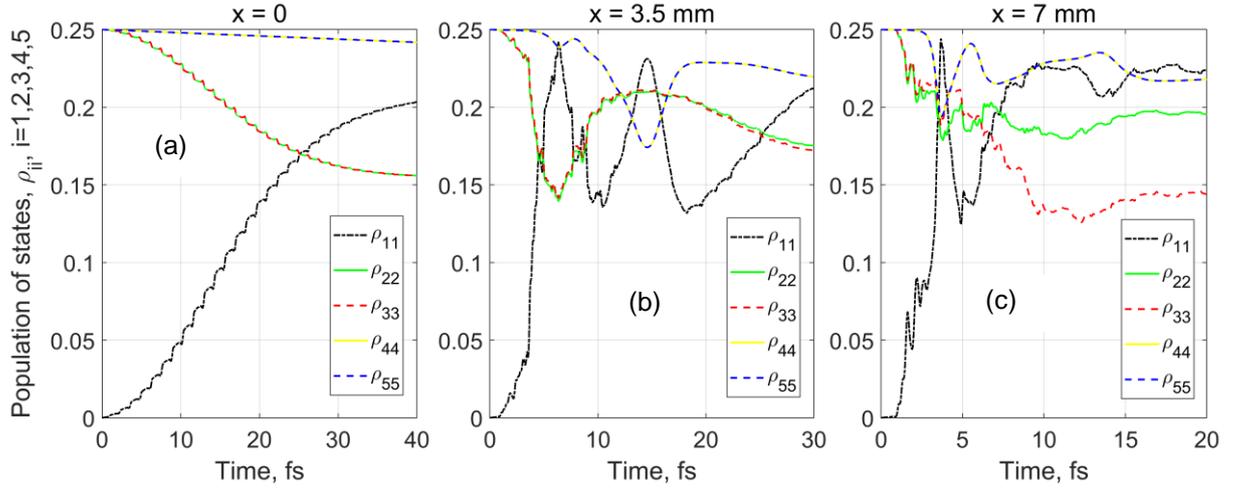

FIG. S9. Time dependencies of populations of the relevant states of the ions: $\rho_{11}$ (black dash-dotted curve), $\rho_{22}$ (green solid curve), $\rho_{33}$ (red dashed curve), $\rho_{44}$ (yellow solid curve), and $\rho_{55}$ (blue dashed curve) for $t_{peak} = 10$ fs, $I_0 = 10$ TW/cm$^2$, and $L = 7$ mm. Panels (a), (b), and (c) correspond to different points within the medium: entrance to the medium, $x=0$ (a), centre of the medium, $x = L/2 = 3.5$ mm (b), and exit from the medium, $x = L = 7$ mm (c).